# NUMERICAL ANALYSIS OF FLEXIBLE ULTRA-THIN CONFORMAL ANTENNA ARRAY

M. Zadeh, A.C. Manzoni
*Department of Electrical Engineering*
*École Polytechnique Fédérale de Lausanne, Lausanne, Switzerland*

## Abstract

We present a flexible ultra-thin conformal patch antenna array operating in the Industrial, Scientific, and Medical (ISM) band (2.4-2.4835 GHz) for wireless sensor nodes. The antenna was designed on a flexible polyimide film (Kapton HN) substrate with the size of 95×18×0.1 $mm^3$ and the number of 7 radiating elements. The antenna exhibits $S_{11} < -10$ dB at the band of interest with broadband behavior. The measured gain is 4.72 dB in vertical polarization at 2.48 GHz. To obtain the maximum gain of the array, distances of the elements in the array are optimized by using the genetic algorithm (GA) in conjunction with the multilevel fast multipole algorithm (MLFMA). The VSWR is less than 2 over 2.408-2.555 GHz, frequency band. The method of moment (MOM) based on the electric field integral equation (EFIE) is applied to analyze the radiation characteristics of the proposed antenna. The resulting matrix equation obtained by the discretization of the EFIE is solved iteratively using the MLFMA. The radiation patterns and return loss obtained from numerical analysis and simulations show a good impedance matching and a gain enhancement of the proposed antenna.

## 1. Introduction

In the recent past, wireless sensor networks (WSNs) have found their way into a wide variety of applications including environmental monitoring, target tracking, space exploration and biomedical applications. WSNs usually consist of a large number of independent devices called sensor nodes. Each sensor node is equipped with integrated sensors and short-range radio communication systems. Traditionally, radio communication systems are equipped with an omnidirectional antenna and a battery that supplies the energy needed by the sensor node to perform the given task. In this situation, majority of the stored power is wasted only because the power is equally radiated in all directions and not concentrated towards the appropriate direction. One way to reduce power consumption and enhance the performance of the sensor nodes is to use high gain or directional antennas.

Numerous efforts have been made to develop high gain antennas for WSNs [1-6]. In [7] authors designed and fabricated an electronically-steerable directional antenna for WSNs called SPIDA. Antenna was primarily designed for localization and concurrent communication, but it can also be used for direction finding applications. SPIDA is composed of a 3 cm-diameter hexagonal sheet made of copper-clad FR-4 and six legs made of 1mm copper wire. An attractive property of the SPIDA is its simplicity. The idea behind SPIDA originally comes from ESPAR [8], an optimized passive array antenna for use in WSNs. The antenna was optimized with respect to gain using the genetic algorithm (GA) and the finite element method (FEM). ESPAR consists of seven passive elements where a single feed monopole is surrounded by six, equally spaced, parasitic monopole elements. The parasitic elements are loaded with varactor diodes to produce various antenna radiation patterns. A fascinating directional antenna is proposed in [9]. The proposed antenna is composed of four planer patch antennas arranged in a box like structure, which is made of RF-4 substrate. The RF signal is dispersed to the four faces by a switch which allows the wireless node to dynamically select which face to use.

Even though the aforementioned works show the performance improvement of the radio communication (high directivity, peak gain, and direction finding) in WSNs, they are rather cumbersome and, moreover, occupies a large volume. These antennas are particularly appealing for stationary or fixed wireless sensor nodes. They do not have the fundamental properties that an antenna should have to be consistent with mobile or nomadic wireless sensor nodes. Properties like flexibility, ultra-thin and low profile which are crucial for mobile WSNs.

In this work a flexible, ultra-thin, low cost conformal patch antenna for WSNs applications in the ISM band is proposed. The antenna was designed on a single layer Kapton® polyimide film substrate featuring a permittivity $\varepsilon_r = 3.5$, a loss tangent tan $\delta \leq 0.004$ and a thickness of 0.1 mm.

To analyze the proposed patch array antenna precisely, we consider all the radiation characteristics by formulating with the electric field integral equation (EFIE). A full wave method of moment (MOM) [10, 11] is used to convert the integral equation into a matrix equation. Discretization of MOM gives rise to a dense matrix of N linear equations and N unknowns which can be solved iteratively [12-15]. The multilevel fast multipole algorithm (MLFMA) have been used to speed up the matrix vector multiplication, in this scheme from $O\left(N^2\right)$ to $O\left(N\ logN\right)$ and reduce memory requirement from $O\left(N^2\right)$ to $O\left(N\ logN\right)$, where $N$ is the number of unknowns [16, 17].

The remainder of this paper is organized as follows: in Section 2 the design of proposed antenna and simulation results are described; later on in Section 3 MLFMA analysis and numerical results are shown and discussed. Finally, conclusions are reported in Section 4.

## 2. Antenna Design

Initially, a flexible, low-profile, ultra-thin and small antenna, having high gain with 0.1 mm thickness to operate over 2.4-2.4835GHz, ISM band is designed.

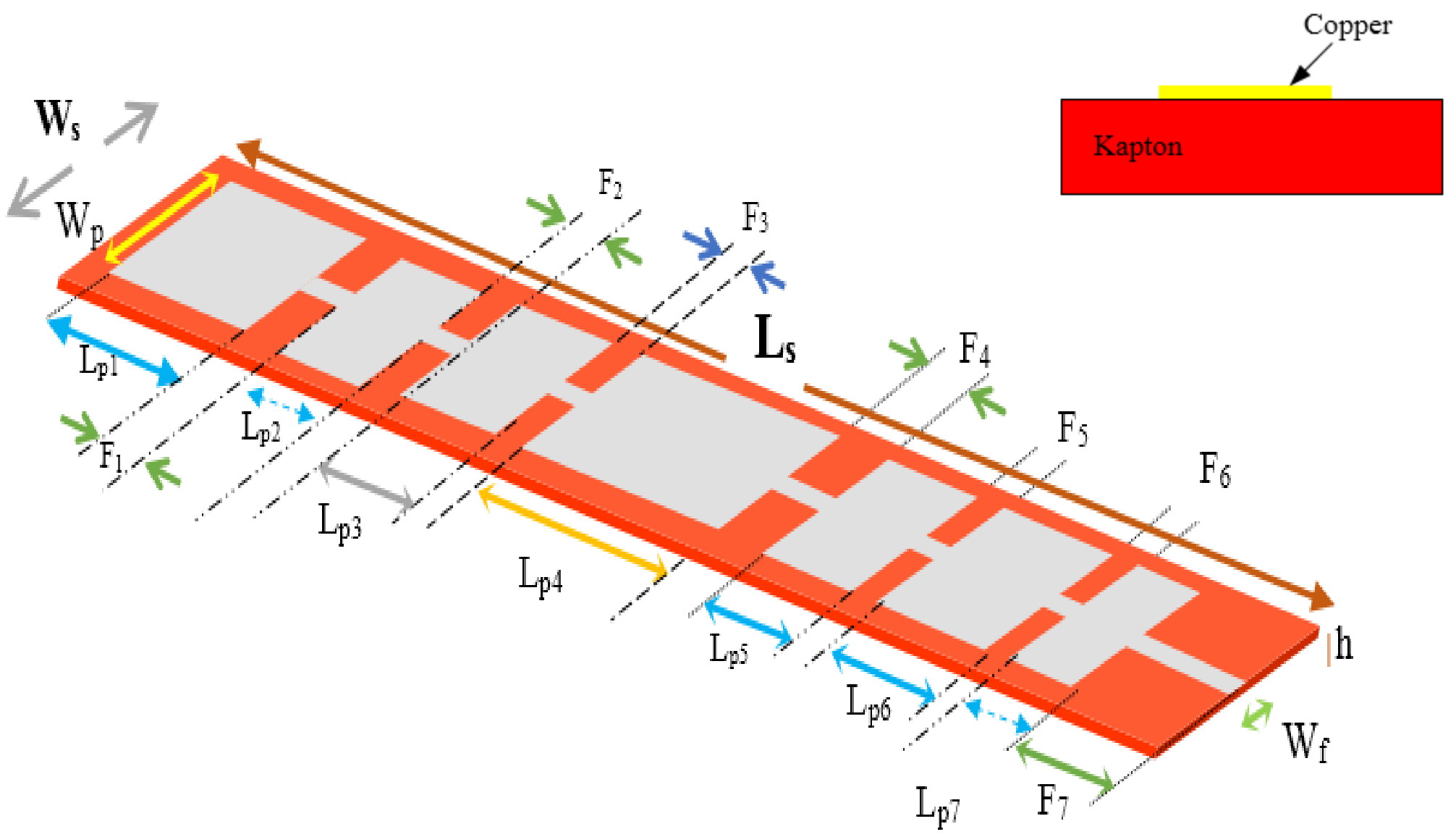

Figure 1: geometry of proposed conformal patch antenna array

The geometry of the conformal patch antenna array introduced in this paper is shown in figure 1. Antenna is designed on the polyimide film with the thickness of h=0.1 mm, length of $L_s$ =95 mm and the width of $W_s$ =18 mm. A polyimide film was chosen because its dielectric constant is not very sensitive to temperature changes. Polyimide film possesses excellent electrical, chemical and physical properties which makes it a suitable choice for WSNs [18]. In order to optimize the antenna array, genetic algorithm (GA) was employed in conjunction with MLFMA. The GA concentrated on improving the horizontal gain of the antenna and antenna structure. Although the authors have applied GA to achieve the desired gain and antenna structure, the design procedure have not been discussed [8]. The optimized distances of the antenna array are shown in Table 1.

Table 1: optimized distances for proposed antenna (unit: mm)

| Parameter | Value | Parameter | Value |
|---|---|---|---|
| $L_s$ | 95 | $L_{p6}$ | 9.5 |
| $W_s$ | 18 | $L_{p7}$ | 3.5 |
| h | 0.1 | $F_1$ | 4.5 |
| $W_f$ | 1 | $F_2$ | 4.5 |
| $W_p$ | 16 | $F_3$ | 3 |
| $L_{p1}$ | 13 | $F_4$ | 5 |
| $L_{p2}$ | 5 | $F_5$ | 1.5 |
| $L_{p3}$ | 8 | $F_6$ | 1.5 |
| $L_{p4}$ | 14 | $F_7$ | 6.5 |
| $L_{p5}$ | 9.5 | | |

The patch antenna array design and optimization have been carried out using CST microwave studio 2014. The effects of polyimide film thickness on the patch antenna reflection coefficient is shown in figure 2.

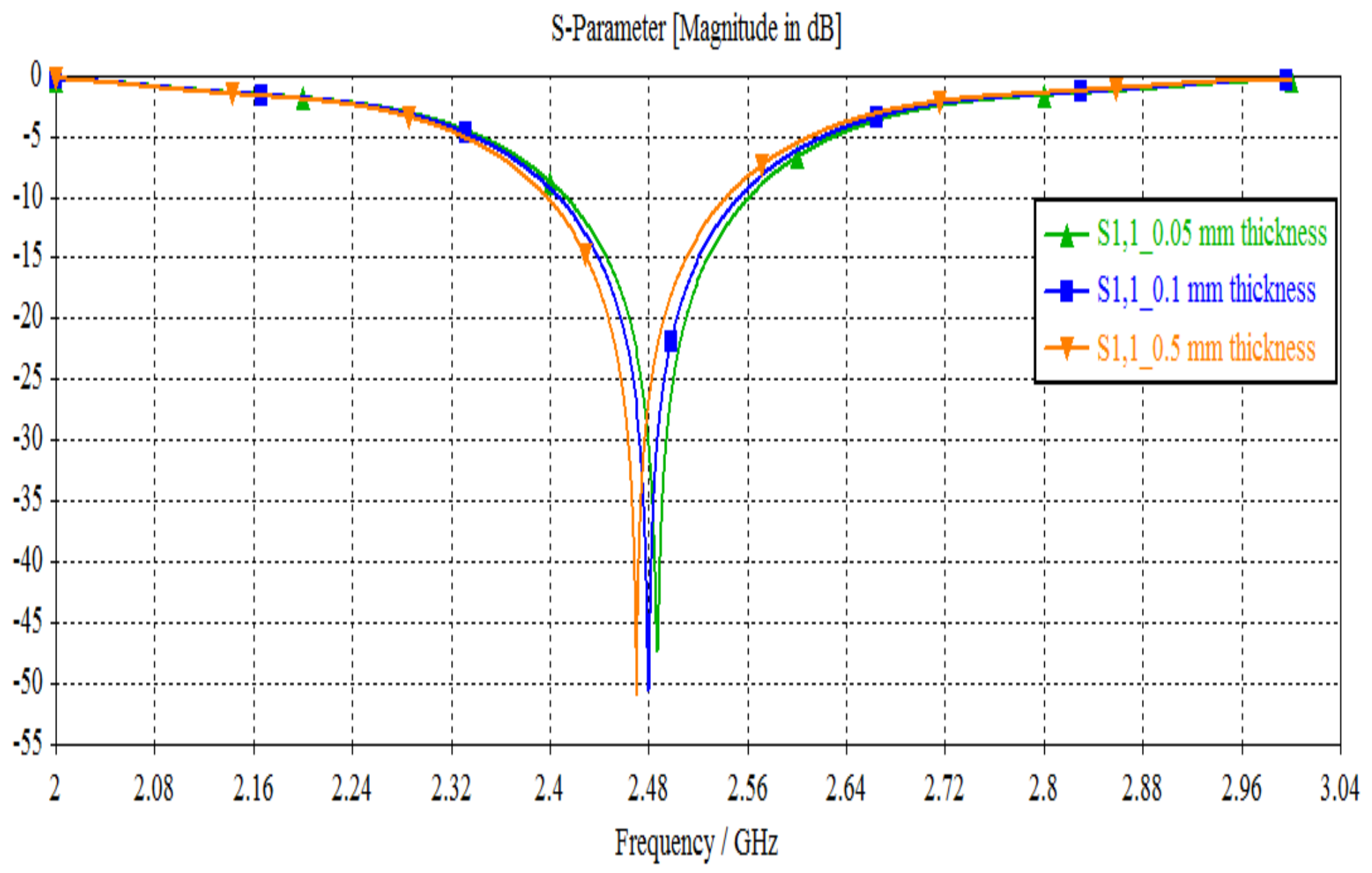

Figure 2: simulated $|S_{11}|$ with the varying of the substrate thickness

As depicted in figure 2, the best substrate thickness (h) for our antenna to work at 2.48GHz is 0.1 mm. While the 0.1 mm polyimide film does resonate at 2.48GHz, its return loss at resonance is ~

-51 dB or 0.0007% power reflected back to the source. It is also observed that the bandwidth of antenna array is about 160 MHz (2.4-2.56GHz).

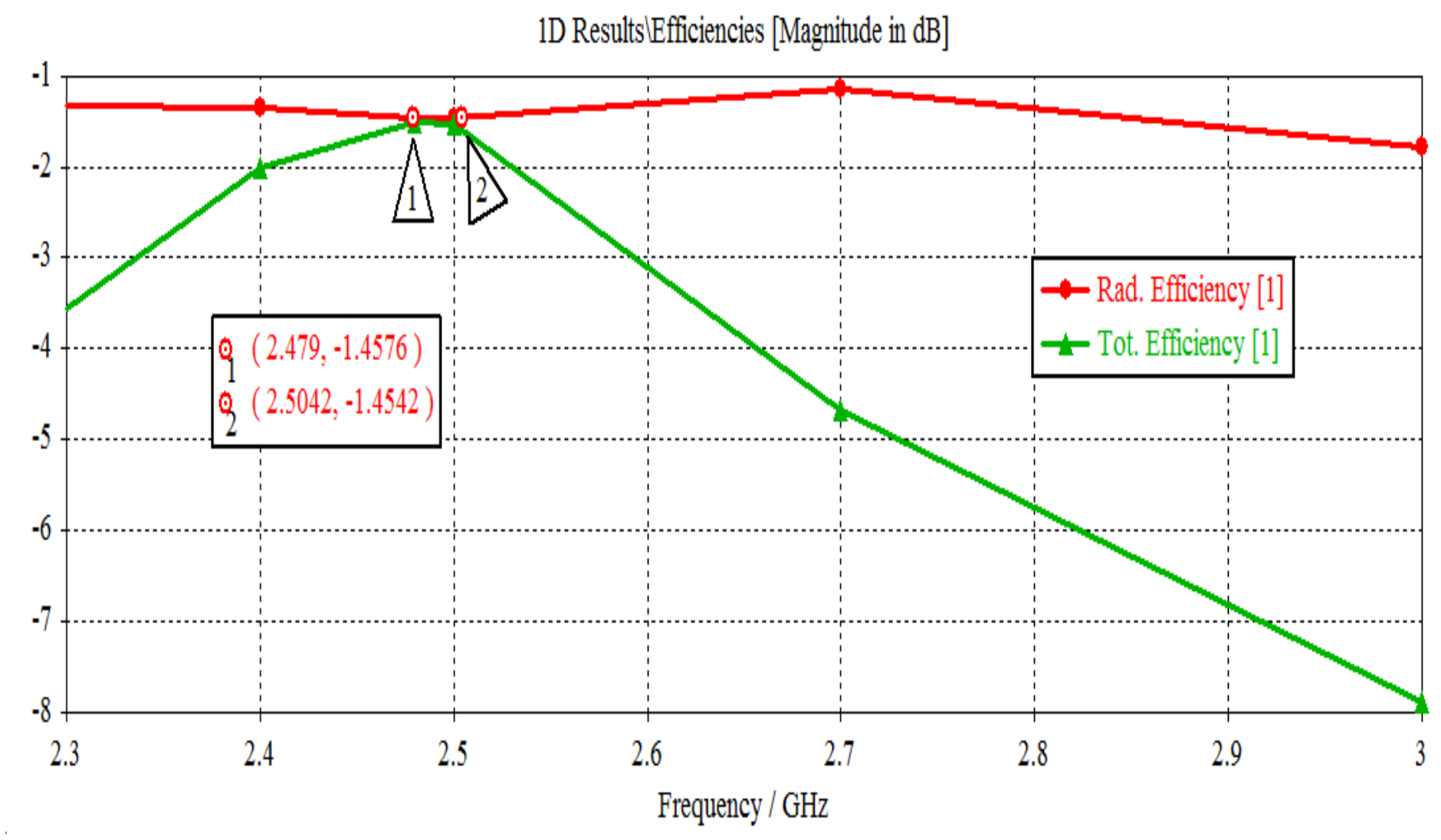

Figure 3: radiation and total efficiency

The radiation efficiency ( $\delta_r$ ) of about -1.45dB or 72% has been achieved at 2.48 GHz.

Using the following equation, which describes the relation between directivity and gain of an antenna, i.e.

$$G(\theta,\varphi)=\delta_r D(\theta,\varphi) \quad (1)$$

We can receive the directivity of 6.55 dB in theory. The voltage standing wave ratio (VSWR) remains below 2:1 for the ISM band. The gain for proposed antenna array is plotted against frequency and is shown in figure 4. As we can see in this figure, antenna gain of 4.72 dB can be obtained at 2.48 GHz.

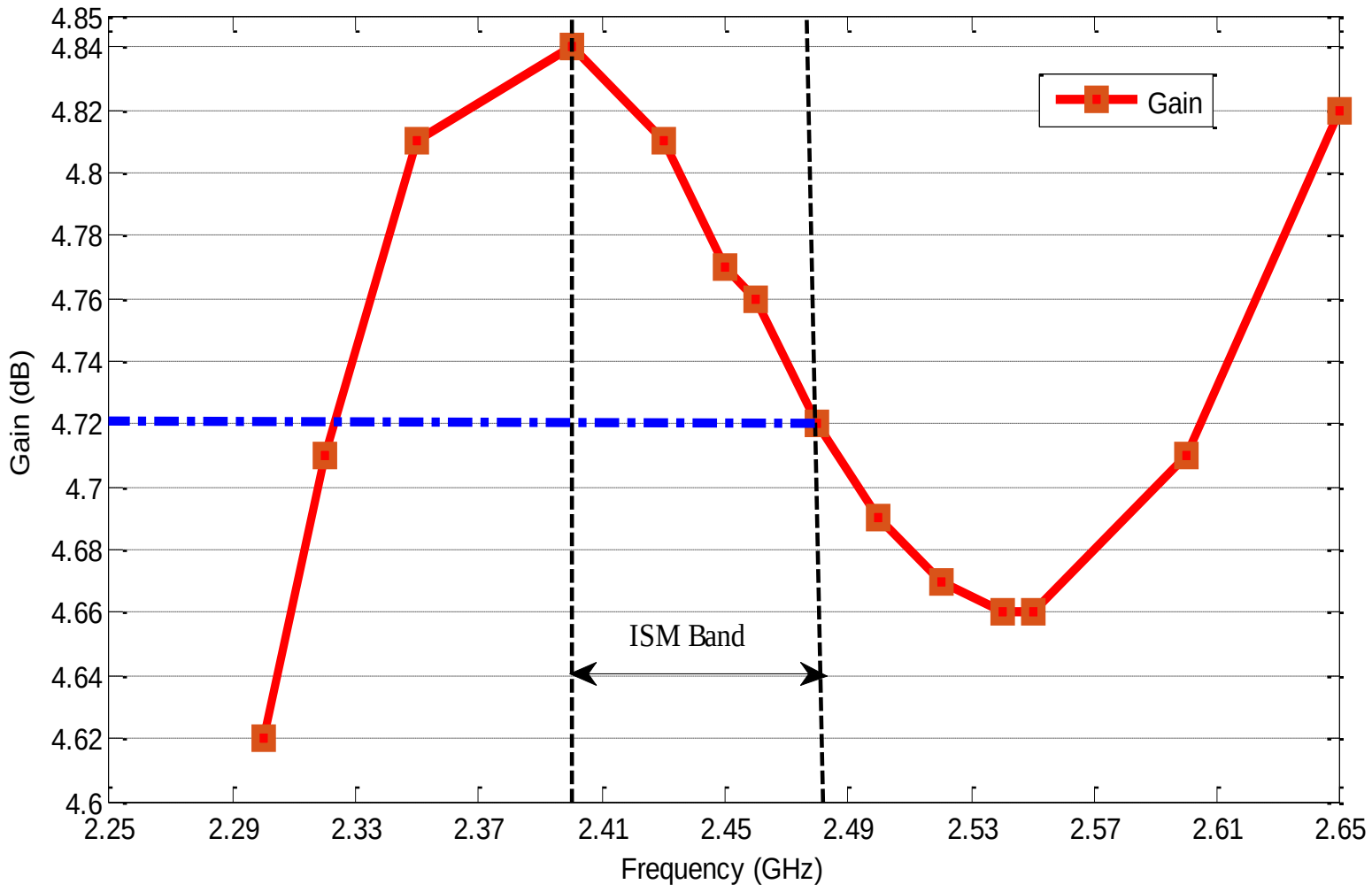

Figure 4: Gain vs. frequency

# 3. Measurement and Analysis

## 3.1. Radiation Characteristics

The far-field radiation patterns of proposed antenna are studied in this subsection. For the brevity, only radiation patterns at 2.4 GHz and 2.48 GHz are presented. For other frequencies (at ISM band), radiation patterns are about the same as those at 2.4 GHz and 2.48 GHz.

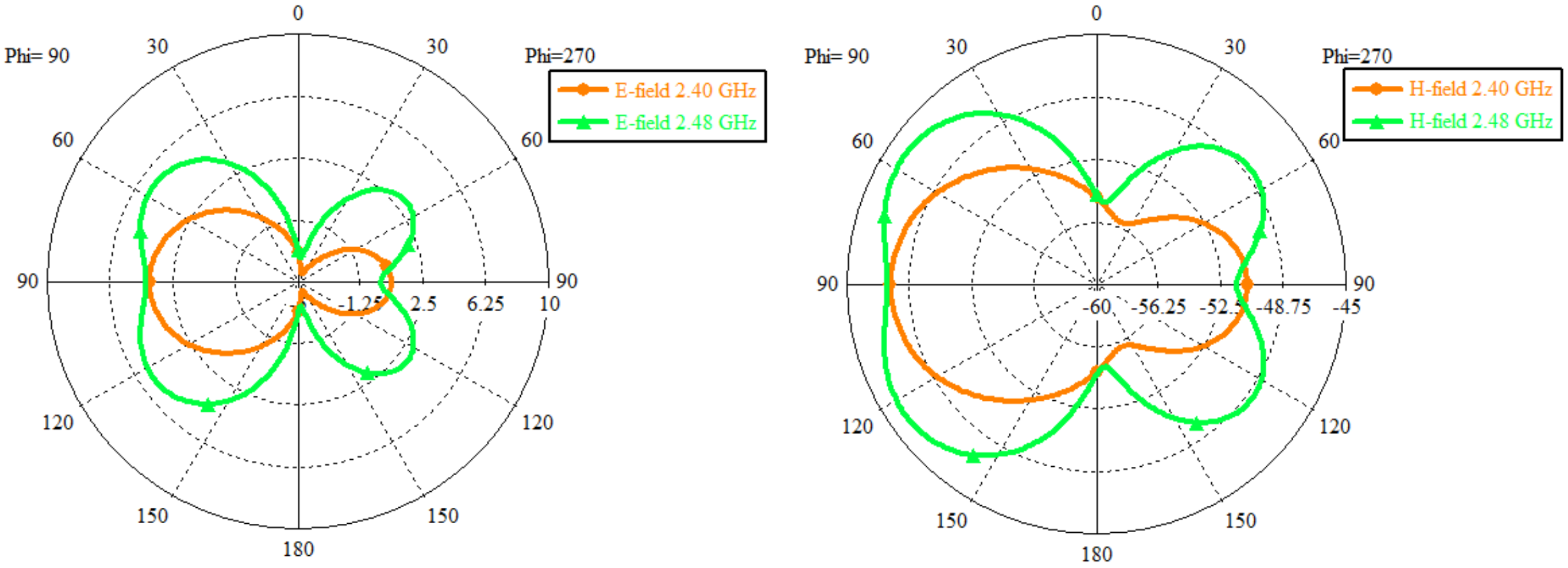

Figure 5: Electric and magnetic radiation pattern **a)** E-field **b)** H-field

Figures 5(a) and (b) show the measured E-plane and H-plane respectively. It is clear that the antenna radiation patterns are nearly directional at both frequencies.

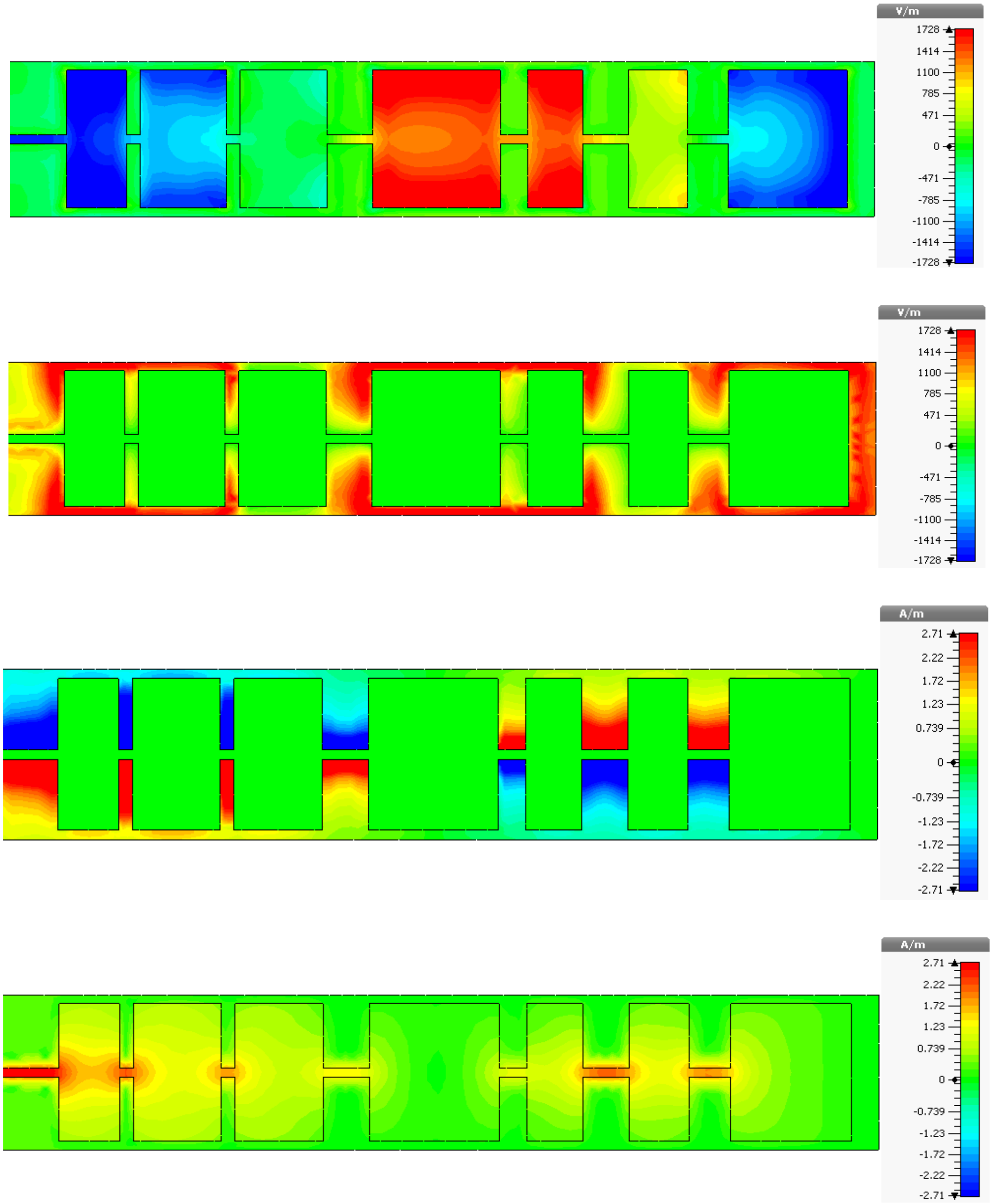

Figure 6: Electric and magnetic field distribution at 2.48 **a)** E-normal **b)** E-tangential **c)** H-normal **d)** H-tangential

Figure 6 shows the simulated normal and tangential ($H_y$) electric and magnetic distributions of proposed antenna at 2.48 GHz, excited by the microstrip patch antenna.

### 3.2. Numerical Results

To further analyze the proposed antenna array, the surface (patch) of the antenna is modelled as a metallic scatterer (perfect electric conductor (PEC)). Then, the scattering and radiation problem is formulated by electric field integral equation (EFIE). The method of moment (MOM) is used to solve the obtained integral equation. Meanwhile, the multilevel fast multipole algorithm (MLFMA) is employed to speed up matrix-vector multiplication and reduce memory requirements and computational complexity [19].

The EFIE can be derived by applying the boundary condition about electric field tangential component. For a PEC (S):

$$\hat{n}.E_{sca}(r) = -\hat{n}.E_{inc}(r) \qquad on\ S \qquad (2)$$

Where the $\hat{n}$ is a normal on the surface S that points outward and r is the observation point which is located on the surface. In the equation, $E_{inc}(r)$ is the incident electric field and $E_{sca}(r)$ is the scattered field, which can be written in terms of electric surface current ( $\mathrm{J_S(r)}$ ) as:

$$E_{sca}(r) = i\omega\mu\int_S dr'\,\bar{\mathrm{G}}(\mathrm{r},\mathrm{r}').\mathrm{J}_S(r') \qquad (3)$$

Where $\eta = \sqrt{\mu/\varepsilon}$ , $\kappa^2 = \omega^2\mu\varepsilon$ is a constant and $\bar{\mathrm{G}}(\mathrm{r},\mathrm{r}')$ is the dyadic Green's function:

$$\bar{\mathrm{G}}(\mathrm{r},\mathrm{r}') = \left[\bar{I} + \frac{\nabla\nabla}{\kappa^2}\right]\frac{e^{ik|r-r'|}}{4\pi|r-r'|} \qquad (4)$$

Then, the EFIE can be formulated by substituting the scattered field $E_{sca}(r)$ into the boundary condition as:

$$\hat{n}.i\omega\langle\bar{\mathrm{G}}(\mathrm{r},\mathrm{r}')\,\mu(r'),\mathrm{J}_S(r')\rangle = \hat{n}.i\omega\int dr'\,\bar{\mathrm{G}}(\mathrm{r},\mathrm{r}')\mu(r').\mathrm{J}_S(r') = -\hat{n}.E_i(r) \qquad (5)$$

The method of moment (MOM) is used to solve the integral equation. MOM finds the matrix representation of EFIE by first expanding the unknown electric surface current in terms of a finite set of expansion functions [20]:

$$J(r) = \sum_{i=1}^{N} f_i(r) I_i = \overline{f}(r).I \tag{6}$$

In the equation, $f_i(r)$ is known as basis or expansion function, while $I_i$ is the unknown coefficient. In order to discretize the surface of the antenna and EFIE, one popular choice is the Rao-Wilton-Glisson (RWG) basis function [21], which discretizes the surface of the antenna into small elements each of which is triangle.

$$f_i(r) = \begin{cases} \dfrac{\rho_i^+}{2\Lambda_i^+}, & r \in T_i^+ \\ -\dfrac{\rho_i^-}{2\Lambda_i^-}, & r \in T_i^- \\ 0, & otherwise \end{cases} \tag{7}$$

Where $\Lambda_i^{\pm}$ is the area of the two adjacent triangles associated with the $i-th$ basis, and $\pm\rho_n^{\pm}(r)$ is the vector from the point r to the vertices. Thus, we have reduced the search for the unknown surface current to the search for a set of unknown coefficients denoted by $I_i$. By substituting (6) and (7) in (5) and multiplying the result by $-\hat{n}\times J_p(r)$, to remove the r, we have:

$$-\int_S drE_i(r).J_p(r) = i\omega\mu\sum_{q=1}^{N} I_q \int_S drJ_p(r).\int_S dr'\overline{G}(r,r').J_q(r') \tag{8}$$

In which $p=1,\ldots,N$. The final matrix-vector form for MLFMA to solve comes from (8) as follows:

$$Z.I=V \qquad (9)$$

Where

$$Z=\left\{Z_{pq}\right\}=i\omega\mu\left\langle J_p,\bar{\mathrm{G}},J_q\right\rangle=i\omega\mu\int_S dr\,\mathrm{J_p(r)}.\int_{\mathrm{S}}\bar{\mathrm{G}}(\mathrm{r,r'}).\mathrm{J_q(r')}dr' \qquad (10)$$

Represents the matrix element, and

$$V=\left\{V_p\right\}=-\left\langle \mathrm{J_p},E_i\right\rangle=-\int_S \mathrm{J_p(r)}.E_i(r)dr \qquad (11)$$

Iterative solution of the N×N matrix equation in (9) via MLFMA gives the coefficients $I_i$ for the expansion in (6). As it is well-known, the MLFMA is one of the most effective numerical methods to solve the integral equation of electromagnetic radiation and scattering [22]. The main idea behind MLFMA, which is based on the factorization of the Green's function, is to substitute element-to-element interactions with group-to-group or cluster-to-cluster interactions. By applying the grouping technique repeatedly, we can develop the MLFMA with a complexity of $N$ per iteration. Comparing with the complexity of applying an iterative solver directly to a classical MOM, which is $O\left(N^2\right)$, the MLFMA apparently is more efficient. The MFMA constructs a tree by putting the scatterer in a large cubic box of edge length $\psi$ and recursively dividing it into eight smaller sub-boxes, until the edge length $\xi_l=\frac{\psi}{2^l}$ at the lowest level $l$ is approximately half a wavelength [23] $O(\log N)$. Vacant boxes are neglected, and only occupied boxes are considered in the tree. At the lowest level each box contains a part of the discretized scatterer, i.e., triangles. In this paper we use:

$$\xi_l = \gamma . {}^{2\pi}\!\!/_{\kappa} \qquad 0.3 \le \gamma \le 0.55 \qquad (11)$$

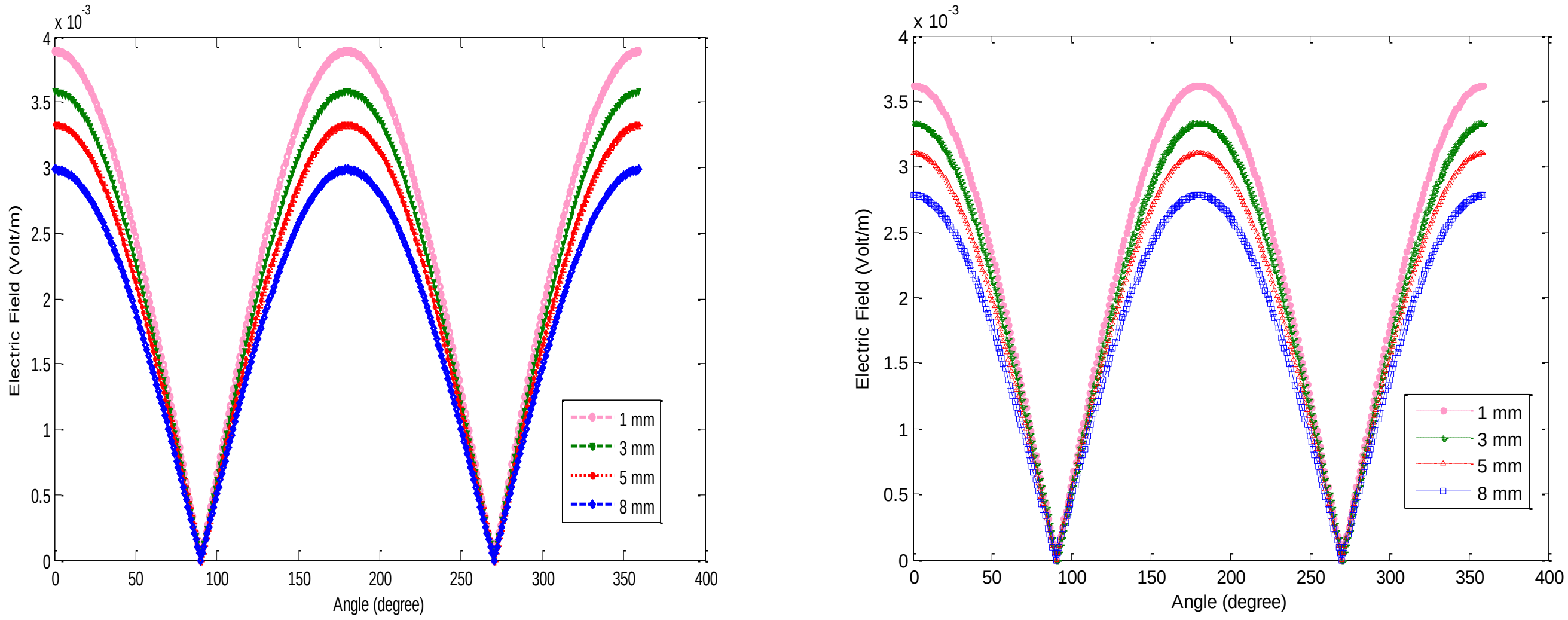

(a) (b)

Figure 7: electric field of the proposed antenna at various frequencies with different mesh sizes via MLFMA **a)** 2.48GHz **b)** 2.40 GHz

Figure 7 presents the electric field strength of the proposed antenna array via 3-level MLFMA in the ISM band with different mesh sizes.

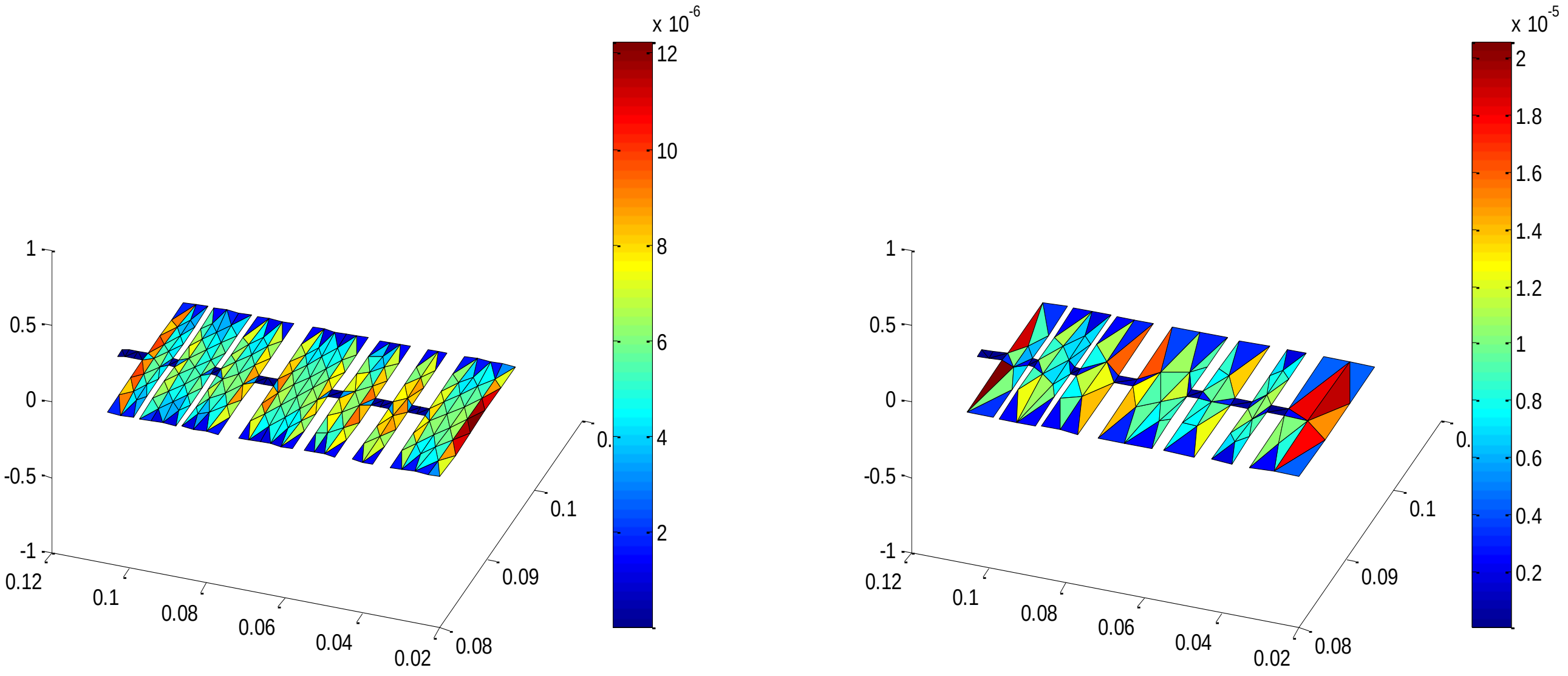

(a) (b)

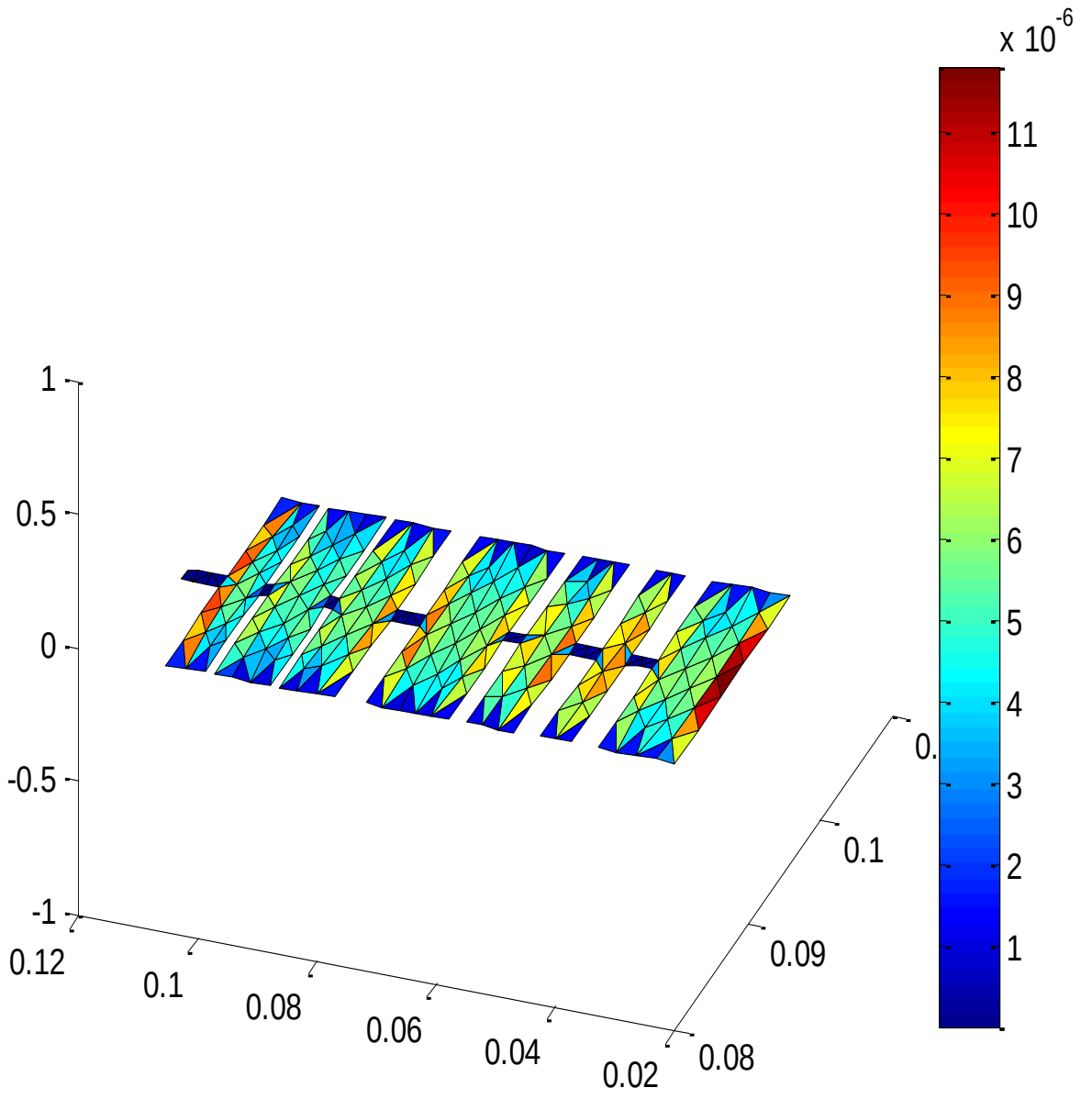

(c)

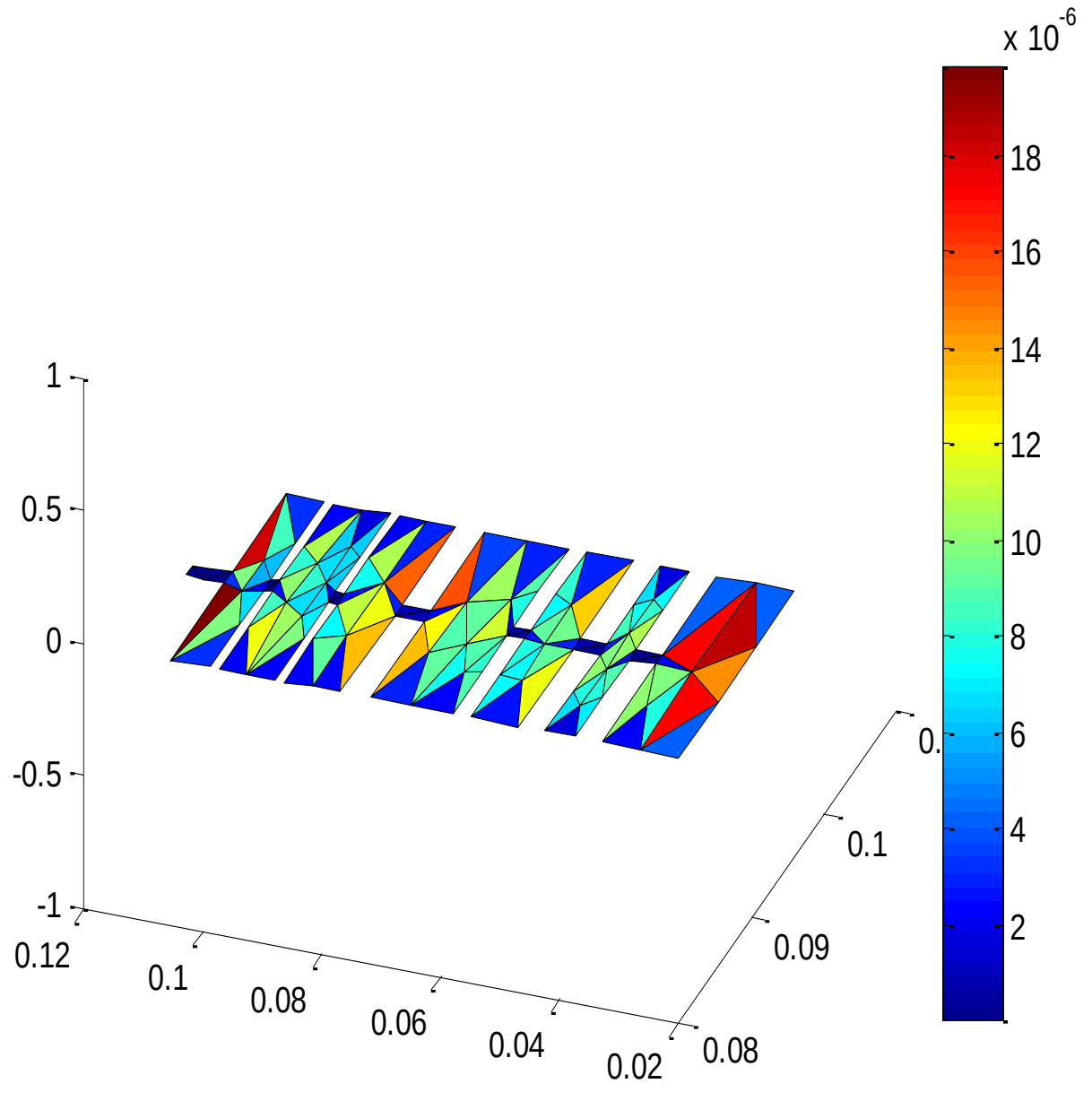

(d)

Figure 8: the surface current distribution **a)** 3 mm mesh, at 2.48 GHz **b)** 6 mm mesh, at 2.48 GHz **c)** 3 mm mesh, at 2.40 GHz **d)** 6 mm mesh, at 2.40 GHz

Eclectic Surface current density on the antenna surface is depicted in figure 8. The excitation is a plane-wave excitation (1 V/m) in the z direction and the transmission line is on left-hand side of the arrays. As we can observe the antenna has approximately the same electric surface current, which makes it suitable for both 2.48 GHz and 2.4 GHz.

## Conclusion

In summary, a novel flexible, ultra-thin conformal patch antenna was presented for operating at 2.4-2.4835 GHz with 4.72 dB gain at 2.48 GHz and bandwidth of 160 MHz (2.4-2.56GHz). Simulation results indicated that 72% radiation efficiency is achieved by using the GA and the polyimide film substrate. To analyze the radiation and scattering characteristics, the problem is formulated by EFIE and it is solved accurately via MOM. The MLFMA is employed to speed up the time and reduce the memory requirements. Numerical and simulation results validate the proposed conformal antenna which can be a good candidate for future mobile wireless sensor nodes.